\documentclass[aps,twocolumn,reprint,preprintnumbers,superscriptaddress,amsmath,amssymb,floatfix]{revtex4-1}
\usepackage{graphicx}% Include figure files
\usepackage{dcolumn}% Align table columns on decimal point
\usepackage{bm}% bold math
\usepackage{amssymb}
\usepackage{epsfig}
\usepackage{epsfig}
\usepackage{amsmath}
\usepackage{amssymb}
\usepackage{amsfonts}
\usepackage{mathptmx}
\usepackage{dcolumn}
\usepackage{eucal}
\usepackage{epstopdf}
\usepackage{bm}
\usepackage{color}
\usepackage{graphicx}
\usepackage{natbib}
\usepackage[colorlinks,linkcolor=blue,citecolor=blue]{hyperref}

\begin{document}
%\title{Efficient temperature-to-frequency converter based on normal metal-ferromagnetic insulator-superconductor junctions}
\title{Superconducting transistors: A boost for quantum computing}
\author{Francesco Giazotto}
\email{francesco.giazotto@sns.it}
\affiliation{NEST Istituto Nanoscienze-CNR  and Scuola Normale Superiore, I-56127 Pisa, Italy}
\begin{abstract}
A niobium titanite nitride-based superconducting nanodevice in which the Josephson critical current can be modulated by a gate voltage - a Cooper-pair transistor - has proven a remarkably long  parity lifetime exceeding one minute at temperatures close to absolute zero.
\end{abstract}
\maketitle

The impressive developments reached so far in nanotechnology  allow routinely the fabrication  of nanosized tunnel junctions, i.e., contacts containing an ultrathin dielectric layer acting as a barrier for charge carriers. In such a case the corresponding junction  capacitance, $C$, can be as low as a few tenths of fF or even below so that the \emph{charging} energy $E_C=e^2/2C$  ($e$ is the electron charge) associated with the accumulation of a single electron at the junction can be of the order of several Kelvins. 
The transport properties of systems based on nanoscale tunnel junctions can therefore be deeply affected by charging effect at sub-Kelvin temperatures and, when this occurs, the systems are are said to be dominated by Coulomb blockade \cite{devoret}.

\begin{figure}[t!]
\includegraphics[width=\columnwidth]{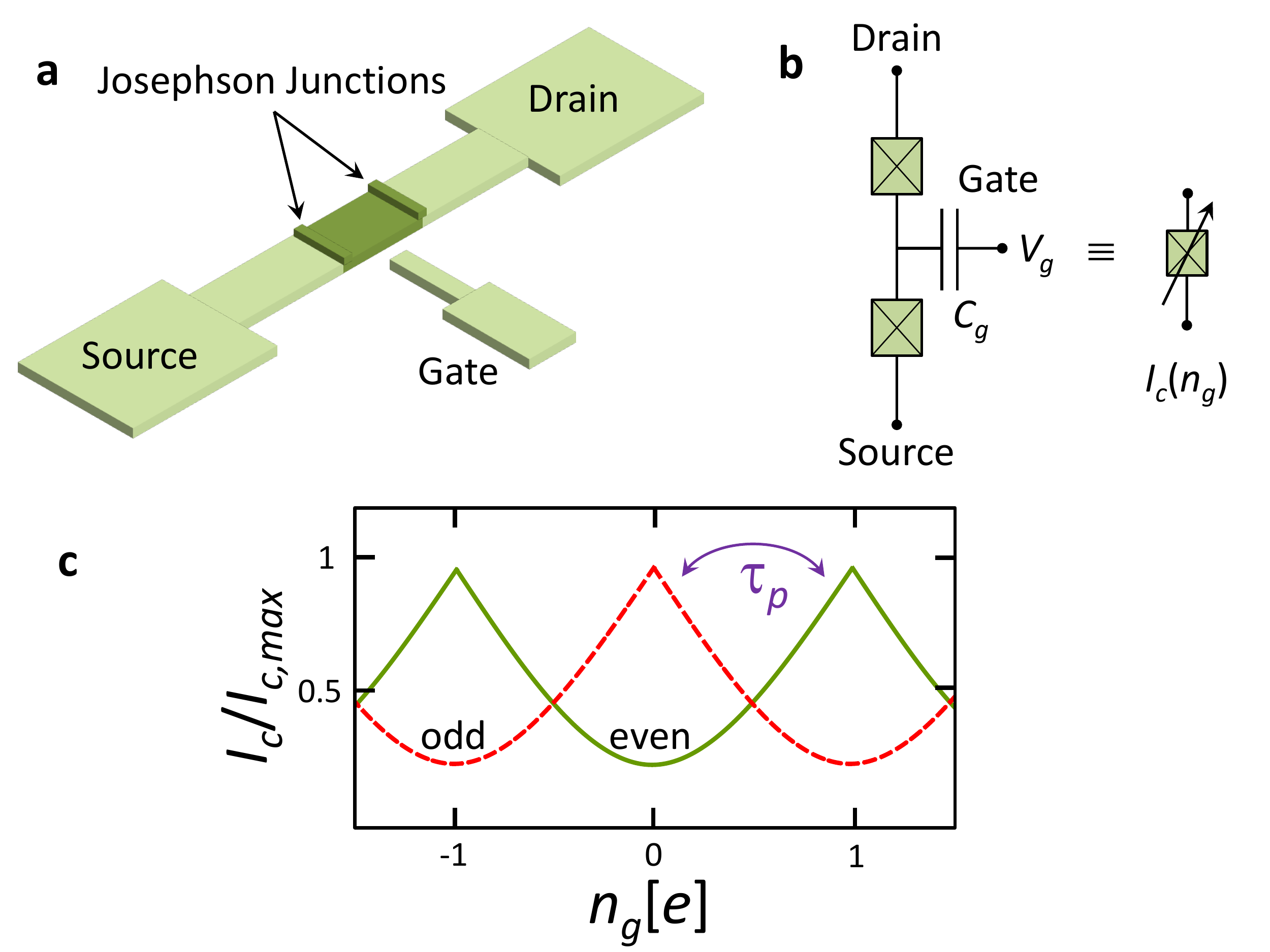}
%\vspace{-8mm}
\caption{\label{fig1} 
\textbf{The Cooper-pair transistor used by van Woerkom and colleagues \cite{vanwoerkom}.} \textbf{a}, The device consists of a superconducting island connected via high-resistance tunnel junctions to source and drain superconducting electrodes. A metallic gate capacitively-coupled to the island allows  modulation of the Josephson supercurrent flowing through the device.
The source and drain leads are made of aluminium, the central island (which is 100 nm thick, 450 nm long and 250 nm wide) is realized with NbTiN, a superconductor with a critical temperature of $\sim 14$ K, and the gate electrode  is made of a titanium/gold bilayer. The CPT tunnel junctions (with area $200\times 250$ nm$^2$) are made of AlO$_x$, and possess normal-state resistance of $\sim 65$ k$\Omega$.
\textbf{b}, The CPT behaves as a Josephson junction whose critical current $I_c$ can be modulated through gate polarization charge, $n_g=C_gV_g/e$, where $C_g$ is the gate capacitance and $V_g$ is the applied voltage.
\textbf{c}, Low temperature qualitative behavior of the CPT critical current $I_c$ versus $n_g$. 
Green and red-dashed lines describe the supercurrent in the \emph{even} and \emph{odd} parity state, respectively, the latter corresponding to a single excess electron on the NbTiN island. 
Each curve is $2e$ periodic in $n_g$ stemming from the Josephson coupling existing in the device.
$\tau_p$ indicates the parity lifetime: Parity may switch on a timescale of $\tau_p$ owing to a single unpaired electron tunneling through the junctions.
$I_{c,max}$ denotes the maximum achievable Josephson critical current.
}
%\vspace{-2mm}
\end{figure}

Within the wide variety of nanodevices governed by Coulomb blockade a special position  is occupied   
by the Cooper-pair transistor (CPT) \cite{nazarov,lafarge}, a fully superconducting three-terminal  device consisting of a mesoscopic island joined to source and drain electrodes via ultrasmall and highly-opaque
tunnel junctions (Fig. \ref{fig1}a). CPTs constitute the basic element in several applications ranging  from metrology and superconducting qubits to ultrasensitive electrometers. 
Due to the smallness of the island, the energetic cost required to add a single Cooper pair to it can largely exceed the operation temperature of the device  (typically below 100 mK) so that the transport properties of the transistor are dependent on the polarization charge presented by a metallic gate electrode capacitively-coupled to it. 
The CPT can therefore be considered as a gate-tunable Josephson junction sensitive to the charge state of its superconducting island (Fig. \ref{fig1}b).    

Although the supercurrent flowing through the device is expected to be $2e$-periodic as a signature of Cooper-pair transport, undesired tunneling of a single unpaired electron through the junctions changes the island charge offset by $e$  thereby modifying its \emph{parity} (Fig. \ref{fig1}c) \cite{tuominen,eiles,joyez}.  
Preserving  parity is crucial for a proper operation of the CPT as its change  is at the origin of undesired supercurrent switching.
The parity lifetime ($\tau_p$) is therefore a fundamental figure of merit and considerable efforts have been made so far to increase it \cite{aumentado}
in particular for improving quantum information processing schemes.
Writing on \emph{Nature Physics}, van Woerkom, Geresdi and Kouwenhoven \cite{vanwoerkom} at Delft University of Technology in the Netherlands report  parity modulation for the first time of a niobium titanite nitride (NbTiN) Cooper-pair  transistor coupled to aluminium (Al) leads  showing an extraordinarily long parity lifetime exceeding one minute at temperatures close to absolute zero.
Remarkably, their CPT can operate in the presence of sizable magnetic fields as large as 150 mT still providing a parity lifetime in excess of 10 ms.

%%%%%%%%FIGURE 1%%%%%%%%%%%%%

The transistor made by van Woerkom and colleagues is based on a small island made of NbTiN (a superconductor with a rather large critical temperature of $\sim 14$ K) connected through  high-resistance tunnel junctions to source and drain aluminum electrodes. 
As in every CPT there are two key-ingredients necessary for its operation. 
The first is the Josephson effect \cite{josephson} which refers to tunneling of Cooper pairs through weakly-coupled superconductors. 
Coulomb blockade, as described above, represents the second ingredient. 
The interplay between the Josephson and Coulomb energy gives rise to the $2e$-periodic modulation of $I_c$ in gate polarization charge shown in Fig. \ref{fig1}c for the \emph{even} (green curve) or \emph{odd} (red-dashed line) parity, which depends on  
the number of excess electrons present on the island.

This idealized picture is drastically modified in a real experiment owing to stochastic
 tunneling of a single quasiparticle 
(a lone electron originating from breaking of a Cooper pair) 
through the CPT junctions. This leads to `poisoning' of the island, 
and reflects into a random changing of the charge parity which yields nasty switchings of the supercurrent, and affects transport periodicity. 
The presence of quasiparticles influences detrimentally the behavior of several superconducting devices, for example, it limits the accuracy of metrological sources or the sensitivity of radiation sensors. 
It is also at the basis of \emph{decoherence} impacting the operation of superconducting qubits. 
Preservation of charge parity is therefore a big concern.  

Among existing superconductors, aluminium has been traditionally the material of choice for CPTs mainly due to the superior characteristics of its oxide for the realization of tunnel junctions. Recent measurements of $\tau_p$ yielded values up to the millisecond range for aluminium-based  devices. 
The exploitation of superconductors with gap (and, accordingly, critical temperature) larger than that of aluminium is believed to be a promising route for improving the CPT performance as this would imply a higher operation temperature, higher working speed, and a reduction of quasiparticle poisoning. 
Yet, a further suppression of poisoning is expected to occur in CPTs possessing the island gap larger than that forming the electrodes, as a larger gap may act as a barrier  for quasiparticles tunneling into the island \cite{aumentado}.

Despite significant efforts no $2e$-periodicity has been reported so far for non aluminium superconductors like, for instance,  in CPTs made of niobium.  
van Woerkom \emph{et al.} \cite{vanwoerkom} have now succeded in realizing the first ever NbTiN-based CPT which demonstrates a dramatic improvement in parity control. 
To put their results into context,  the measured one minute parity lifetime roughly translates  into a single-electron event occurring in their device only once every $\sim 10^{11}$ Cooper pairs have tunneled through the CPT junctions which testifies the smallness of probability of parity switches.
In addition,  the demonstration of parity robustness even in the presence of fairly large magnetic fields is a pivotal asset in light of inducing topological superconductivity in spin-orbit-coupled semiconductor nanowires \cite{mourik}.

It is very likely that the results of van Woerkom and collaborators will have strong impact on the design of improved quantum information processing schemes as well as on parity control and for the readout of  Majorana bound states, for which NbTiN has become the favorite superconductor \cite{mourik}.

\end{document}